# New Exact Solutions for QCD

By Dr. Scott Chapman
scottc@cal.berkeley.edu

**Abstract**
A new class of exact quantum solutions of QCD is presented. These solutions have negative energy and are stable to all fluctuations. The lowest-energy solution state is explicitly constructed and proposed as a candidate for the QCD vacuum. This vacuum exhibits confinement in the sense that any colored quark configurations require infinite energy to create. In addition to the vacuum, other solutions include hadrons whose mass and radius are self-consistently determined as local minima of the energy.

**Introduction**

A considerable amount of work has been done in attempting to analytically derive quark confinement mechanisms from QCD. The simplest approach is to assume that the vacuum has some constant, classical background gluon electric or magnetic field that produces negative energy density and confines color much the same way that magnetic flux is confined in a superconductor [1,2]. A well-known problem with this approach is that since the gluon field is a vector field, the existence of a constant, classical background field implies that the vacuum has some preferred direction in spacetime. This problem can be avoided by assuming a dilute gas of instantons rather than a constant background field [3], but these models are not stable to large fluctuations [4]. Recently, other approaches have been proposed, such as modifying the QCD action functional [5].

In this paper, a new approach is proposed that is a variation on the original background gluon field approach. Instead of assuming a constant, classical background gluon field, one assumes a background gluon field operator made up of quantized quark creation and destruction operators. The functional form of the background field is then determined by the requirement that it leads to exact extrema of the quantized QCD action. The quantum solution states that form these extrema are explicitly constructed, and the lowest-energy "vacuum" state is determined. In contrast to constant, classical background field approaches, it is shown that the vacuum expectation value of the background gluon field vanishes. As a result, the presence of a vector background field does not imply a preferred reference frame for the vacuum.

The concept of a gauge field being made up of fermion operators is not new. In fact it is a well-known result that in QED in the Coulomb gauge, the equations of motion imply that the temporal photon field $A_0$ is equal to a spatial integral of quantized fermion fields [6]. This paper simply extends that idea in a Lorentz-covariant way so that the gluon field $\overline{A}_\mu^a$ that solves the equations of motion is equivalent to a spacetime integral of quark fields.

The approach of this paper can only be used for QCD and not for QED or the electroweak interaction. It will be shown that the reason this approach cannot be used for QED is because it only works for groups that have traceless group generators. It will also be shown that the "diagonal" background field approach used in this paper only works when fermion masses are independent of group indices. Since quark masses are independent of color, this restriction is met for QCD. However, since lepton and quark masses in SU(2) doublets are not independent of their electroweak SU(2) index, the condition is not met for the electroweak interaction.

The outline of the paper is as follows: In the first section, the theory is defined, and the quantum equations of motion in the presence of an operator background field are derived. In the second section, the functional form of the background gluon field is specified, and exact quantum solution states are explicitly constructed out of quark creation and destruction operators. In the third section, the QCD Hamiltonian is derived, energies of the solution states are found, and the lowest-energy vacuum state is identified and shown to be stable to all fluctuations. In the fourth section, it is shown that the vacuum permits finite-energy hadron solutions whose size and mass are self-consistently determined by forming local minima of energy. Confinement of the theory is also demonstrated by showing that any quark configuration that has overall color within the quantization volume would take infinite energy to create.

The most important equations in the paper are the equations of motion in (1.10) – (1.12), the definition of the background field in (2.1)-(2.6), the quantum relation (2.9) that allows solution to the equations of motion, the explicit definition of the vacuum state in (3.12), (2.21), (2.13), and the self-consistently determined mass of a "cubical" baryon in (4.11).

## 1. Definition of the theory

The following action is considered:
$$S = \int d^4x \left[ \overline{\Psi}(i\slashed{\partial} - M + g\slashed{A})\Psi - \tfrac{1}{4} F^a_{\mu\nu} F^{\mu\nu a} \right], \tag{1.1}$$
where $A^a_\mu$ are gluon fields, $\Psi$ are quark (and anti-quark) fields in the fundamental representation, and $M$ is a quark mass matrix in color space. For QCD, the fermion mass matrix is independent of color, so it takes the form
$$M_{\alpha\beta} = m\delta_{\alpha\beta}, \tag{1.2}$$
where $\alpha$ and $\beta$ are indices in the fundamental triplet representation of color SU(3). Standard notation is used in which $\slashed{A} = A^a_\mu \gamma^\mu T^a$ and $T^a = \tfrac{1}{2}\lambda^a$ are half the Gell-Mann matrices. Quantization of the theory will be performed using the canonical method rather than the path integral method. As a result, one can think of the fields in (1.1) as operators rather than as classical fields.

The purpose of this paper is to present a new class of solution states $|\xi\rangle$ that create non-trivial extrema of the quantized action. In other words, new solutions are sought to the equation:
$$\delta\langle\xi'|S|\xi\rangle = 0, \tag{1.3}$$
where $|\xi'\rangle$ may be the same state as $|\xi\rangle$ or some other state in the class of solutions that will be defined shortly. Throughout the paper, $|\xi\rangle$ and $|\xi'\rangle$ will be used only to refer to states that satisfy (1.3).

In the next section, it will be shown that the equations of motion from (1.3) involving quark operators decouple from those involving gluon-fluctuation operators. Consequently, the quark operators in $|\xi\rangle$ can be taken to commute with the gluon-fluctuation operators in $|\xi\rangle$. As a result, these sectors of the Hilbert space can be treated independently, and a general quantum solution state can be written

$$|\xi\rangle = \sum_{nm} c_{nm} |\xi_\Psi^{(n)}\rangle |\xi_A^{(m)}\rangle. \qquad (1.4)$$

Here $|c_{nm}|^2 = 1$, $|\xi_\Psi^{(n)}\rangle$ represents an orthonormal basis of the quark sector of the restricted class of solution states, and $|\xi_A^{(m)}\rangle$ represents an orthonormal basis of gluon-fluctuation sector.

Let us evaluate the contribution to (1.3) from the variation of the operator field $\overline{\Psi}(x)$. In canonical quantization, $\overline{\Psi}(x)$ is made up of every possible positive- and negative-energy quark creation operator allowed by the quantization. The actual value taken by the field at any point in spacetime is determined by the quantum states $|\xi\rangle$ and $\langle\xi'|$ that act on it. Therefore, a complete variation of a matrix element of operator fields must include not only a variation of the functional form of the field $\delta\overline{\Psi}(x)$, but also the variations $\delta|\xi\rangle$ or $\delta\langle\xi'|$. As a result, one finds:

$$\delta_{\overline{\Psi}}\langle\xi'|S|\xi\rangle = \left[ \sum_{nm} (c'_{nm})^* \langle\xi_A^{(m)}| \left(\delta\langle\xi_\Psi^{(n)}|\right) \int d^4x \, \delta\overline{\Psi}(i\partial - M + g\overline{A})\Psi |\xi\rangle \right.$$
$$\left. + \langle\xi'| \int d^4x \, \delta\overline{\Psi}(i\partial - M + g\overline{A})\Psi \sum_{nm} c_{nm} \left(\delta|\xi_\Psi^{(n)}\rangle\right) |\xi_A^{(m)}\rangle \right] = 0. \qquad (1.5)$$

In the above relations, since quark and gluon-fluctuation operators commute, variations of $\overline{\Psi}(x)$ only result in variations of the quark sectors of the states, not variations in the gluon fluctuation sectors of the states. It should be noted that no restriction is placed on the variations of these quark states. As a result, (1.5) ensures that the extremum condition (1.3) is met for any variations of $\overline{\Psi}(x)$, including variations to quark states outside of the class of solution states.

It has been noted that a quantized quark field already has every possible variation included in it; its actual functional form is determined by the quantum states on which it acts. As a result, one can make the following replacement in (1.5):

$$\delta\overline{\Psi}(x) = \overline{\Psi}(x). \qquad (1.6)$$

One can also determine the equation analogous to (1.5) that follows by making variations to the quark field $\delta\Psi(x)$. After making a replacement analogous to (1.6), one finds that the quantum equation of motion for $\Psi(x)$ is identical to that for $\overline{\Psi}(x)$; they are both equal to (1.5) using the replacement (1.6).

It should be noted that $\overline{A}_\mu^a$ is used in (1.5) in place of $A_\mu^a$. The reason for this is that in order to find solution states that cause the variation of (1.5) to vanish, it is necessary to put restrictions on the functional form of the gluon field operator. Thus, $\overline{A}_\mu^a$ is the notation for the restricted form of the gluon field that solves (1.5). Since all of the operators other than $\overline{A}_\mu^a$ in (1.5) are quark operators, a reasonable ansatz for the extremum gluon field is for it also to be made up of quark operators:

$$\overline{A}_\mu^a = \overline{A}_\mu^a(\Psi, \overline{\Psi}). \qquad (1.7)$$

It will be shown that this ansatz does indeed lead to exact solutions to (1.5), and more generally (1.3). Because of the functional form of (1.7), the extremum gluon field acts on the quark states $|\xi_\Psi^{(n)}\rangle$ and not on the gluon fluctuation states $|\xi_A^{(m)}\rangle$.

Since the restriction (1.7) is placed on the functional form of the extremum gluon field, variations around this gluon field may take different functional forms. In order to consider all possible fluctuations around this extremum gluon field, the gluon-fluctuation fields

$$\hat{A}_\mu^a(x) \equiv \delta A_\mu^a(x) \tag{1.8}$$

are quantized independently. It is the quantized operators inside $\hat{A}_\mu^a$ that appear in the gluon-fluctuation states $|\xi_A^{(m)}\rangle$.

Keeping this in mind, a general variation of the gluon field around the extremum of the action is given by

$$\delta_A \langle \xi'|S|\xi\rangle = \left[ \sum_{nm} (c'_{nm})^* \left( \delta \langle \xi_A^{(m)}| \right) \langle \xi_\Psi^{(n)} | \int d^4x \, \delta A_\nu^a \left( -\overline{D}_\mu^{ab} \overline{F}^{\mu\nu b} + g\overline{\Psi}\gamma^\nu T^a \Psi \right) |\xi\rangle + \right.$$

$$\left. + \langle \xi| \int d^4x \, \delta A_\nu^a \left( -\overline{D}_\mu^{ab} \overline{F}^{\mu\nu b} + g\overline{\Psi}\gamma^\nu T^a \Psi \right) \sum_{nm} c_{nm} |\xi_\Psi^{(n)}\rangle \delta |\xi_A^{(m)}\rangle \right] = 0, \tag{1.9}$$

where $\overline{D}_\mu^{ab} = \partial_\mu \delta^{ab} + g\varepsilon^{abc}\overline{A}_\mu^c$. Whereas the variations of (1.5) were only in the quark sector of the states, the variations of (1.9) are only in the gluon-fluctuation sector of the states.

Equations (1.5) and (1.9) are quite complicated. Consider the following simpler "equations of motion":

$$\int d^4x \, \overline{\Psi}\left(i\overline{\partial} - M + g\overline{A}\right)\Psi |\xi_\Psi^{(n)}\rangle = 0 \tag{1.10}$$

$$\langle \xi_\Psi^{(n)}| \int d^4x \, \overline{\Psi}\left(i\overline{\partial} - M + g\overline{A}\right)\Psi = 0 \tag{1.11}$$

$$\langle \xi_\Psi^{(n)}| \int d^4x \, \hat{A}_\nu^a \left( -\overline{D}_\mu^{ab} \overline{F}^{\mu\nu b} + g\overline{\Psi}\gamma^\nu T^a \Psi \right) |\xi_\Psi^{(m)}\rangle = 0, \tag{1.12}$$

Using (1.4) – (1.9), it is easily seen that quark states that satisfy the above equations will also satisfy (1.3), thus forming nontrivial extrema of the quantized action. It should be noted that solutions to the above equations only place restrictions on allowed quark states and no restrictions on allowed gluon-fluctuation states. In the next section, the form of $\overline{A}_\mu^a$ and of several classes of quark solution states will be explicitly constructed.

Before moving on, it is useful to note that the extremum gluon field $\overline{A}_\mu^a$ can also be thought of as a background gluon field. The fact that the background field considered here is an operator rather than a classical field allows it to take a non-trivial form without picking out a preferred direction in spacetime for the vacuum. In particular, it will be shown that the background field satisfies

$$\langle 0|\overline{A}_\mu^a|0\rangle = 0, \tag{1.13}$$

where $|0\rangle$ is the vacuum of the theory. It should be noted that for a classical background field, (1.13) would mean that the background field trivially vanishes. However since the background field considered here is an operator, it can satisfy (1.13) while still satisfying a non-trivial relation such as $\langle 0|\overline{F}_{\mu\nu}^a \overline{F}^{\mu\nu a}|0\rangle \neq 0$.

## 2. The background gluon field and solution states

The proposed background gluon field is equivalent to the following:

$$g\bar{A}_\mu^a(x) \cong -\int d^4 y\, \theta(u \cdot (x-y)) \partial_\mu \overline{\Psi}(y) M \overline{T}^a \Psi(y) \qquad (2.1)$$

where $u$ is the 4-velocity of some as-yet-undetermined reference frame and the non-vanishing components of the color matrix $\overline{T}^a$ are defined by:

$$\overline{T}^3 = \tfrac{3}{2}\lambda^3$$
$$\overline{T}^8 = \tfrac{3}{2}\lambda^8, \qquad (2.2)$$

so that $\overline{T}^a \overline{T}^a = \tfrac{3}{4}(\lambda^3\lambda^3 + \lambda^8\lambda^8) = 1$. The background field of (2.1) manifestly transforms as a Lorentz 4-vector. Although the background field is dependent on an embedded reference frame $u$, it will be shown that the equations of motion and proposed vacuum state are both independent of $u$. As a result, one can choose any value of $u$ that is convenient when performing calculations. It is also interesting that the background field is proportional to the quark mass matrix. It will be shown below that this attribute allows the background gluon field to satisfy an important commutation relation (2.9) that is needed to solve the equations of motion.

It was stated that the background field is "equivalent" to (2.1). The actual form of the background field that will be used in this paper removes temporal nonlocality from (2.1), but is equivalent to (2.1) after using the equations of motion for quark fields and ignoring a surface term. To present the actual definition of the background gluon field, it is necessary to define a quark "potential" field $\bar{\chi}$ via

$$\overline{\Psi} = -i\bar{\chi}\overleftarrow{\partial} = -i\partial_\nu \bar{\chi}\gamma^\nu . \qquad (2.3)$$

Using this field, the background field of (2.1) may be rewritten:

$$g\bar{A}_\mu^a(x) \cong i\int d^4 y\, \theta(u\cdot(x-y))[\partial_\nu(\partial_\mu \bar{\chi}(y)\gamma^\nu M\overline{T}^a\Psi(y)) - \partial_\mu \bar{\chi}(y) M\overline{T}^a \partial\Psi(y)] . \qquad (2.4)$$

Temporal nonlocality of the background field can be removed by assuming that the quark fields satisfy the following "massless" equation

$$\partial\Psi = 0, \qquad (2.5)$$

so that the second term of (2.4) vanishes. It will be shown shortly that this assumption is consistent with solving the equations of motion (1.10) – (1.12). By also assuming that fermion fields vanish at spacelike infinity (relative to $u$), only the temporal part of the derivative in the first term survives, and $\partial_\nu \to u\cdot\partial$. A partial integration of this remaining derivative transforms the theta function in (2.4) into delta functions at $x$ and at temporal negative infinity.

The resulting quantity will be taken to be the definition of the background gluon field:

$$g\bar{A}_\mu^a(x) \equiv i\int d^4 y\, \delta(u\cdot(x-y)) \partial_\mu \bar{\chi}(y) u\cdot\gamma M\overline{T}^a\Psi(y). \qquad (2.6)$$

Due to the fact that (2.1) is a Lorentz 4-vector, the above expression is also a Lorentz 4-vector. It is not immediately apparent why (2.6) is a better choice for the definition of the background field than the more elegant expression in (2.1). The fact that they are equivalent within the context of equation (2.5) and an x-independent surface term means that they both have the same effect in the equations of motion (1.10) – (1.12).

The advantage to (2.6) comes in the derivation of the Hamiltonian. In calculating the Hamiltonian, it is convenient to choose a reference frame $u = (1,0,0,0)$, and use the expression

For $u = (1,0,0,0)$: $\quad g\overline{A}_\mu^a(t) = i\int d^3y \partial_\mu \overline{\chi}(t,\vec{y})\gamma^0 M\overline{T}^a \Psi(t,\vec{y})$. (2.7)

With this value of $u$, henceforward referred to as the "Hamiltonian frame", all fields in the Lagrangian have the same time coordinate, so standard temporally local canonical techniques can be used to derive the Hamiltonian. If one were to take (2.1) as the definition of the background field rather than (2.6), then one would have to use generalized nonlocal techniques to derive the Hamiltonian, since field equations such as (2.5) cannot be applied to Lagrangians before derivation of canonical momenta.

As an aside, it would be interesting to see if the results of this paper could be reproduced by taking (2.1) as the definition of the background field, then using nonlocal techniques such as those presented in [7] to derive the Hamiltonian. A very interesting result of taking (2.1) as the definition of the background field is that it would impose an "arrow" to time. Because of the theta function, the background field of (2.1) at any point of spacetime is dependent on every point in its past light cone as well as other points outside of its future light cone. Despite the philosophical attraction of this nonlocal approach, nonlocal techniques are outside the scope of this paper, so for now (2.6) will be taken as the definition of the background field.

The next step is one that is only possible in a quantized theory and does not have a natural equivalent in a classical field theory. The quark fields are quantized such that

$$\{\Psi^+(t,\vec{x}), \Psi(t,\vec{y})\} = \delta^3(\vec{x}-\vec{y})$$
$$\{\Psi^+(t,\vec{x}), \partial_\mu \chi^+(t,\vec{y})\} = 0. \quad (2.8)$$

That being the case, the background gluon field defined in (2.6) satisfies the following commutation relation:

$$\int d^4x [\overline{\Psi}, g\overline{A}_\mu^a] T^a \gamma^\mu \Psi = \int d^4x \overline{\Psi} M \Psi. \quad (2.9)$$

Equation (2.9) is most easily derived by using (2.7) and (2.8) in the "Hamiltonian frame" $u = (1,0,0,0)$. Since both sides of (2.9) are Lorentz invariants, there is no loss of generality by performing the calculation in the Hamiltonian frame.

Normally in equations of motion, expansions of the fields are performed such that the kinetic term $\overline{\Psi}i\partial\Psi$ cancels the mass term $\overline{\Psi}M\Psi$. However in the case considered here, the "massless" field equation $\partial\Psi = 0$ is needed to ensure that (2.6) is equivalent to (2.1), so that the background field correctly transforms as a Lorentz 4-vector. As a consequence, the kinetic term of the equation of motion vanishes independently and cannot cancel the mass term. Equation (2.9) is important due to the fact that it allows the commutator part of the quark-gluon interaction term to cancel the quark mass term in the equation of motion (1.10).

It is now useful to rewrite equation of motion (1.10) in light of equation (2.9):

$$\int d^4x \left(\overline{\Psi}i\partial\Psi - \overline{\Psi}M\Psi + [\overline{\Psi}, g\overline{A}_\mu^a]T^a\gamma^\mu\Psi + g\overline{A}_\mu^a J^{\mu a}\right)\xi_\Psi^{(n)}\rangle = 0, \quad (2.10)$$

where the colored quark current is given by

$$J^{\mu a} = \overline{\Psi}T^a\gamma^\mu\Psi. \quad (2.11)$$

In (2.10), the first term vanishes due to (2.5) and the next two terms cancel due to (2.9). To completely solve the equation, the last term must vanish independently. This can be accomplished by constructing quark solution states $\left|\xi_\Psi^{(n)}\right\rangle$ that satisfy:

$$\int d^4x g \overline{A}_\mu^a J^{\mu a} \left|\xi_\Psi^{(n)}\right\rangle = \left\langle\xi_\Psi^{(n)}\right| \int d^4x g \overline{A}_\mu^a J^{\mu a} = 0, \tag{2.12}$$

where the second relation above is needed to solve the conjugate equation of motion (1.11). It should be noted that (2.12) is a Lorentz invariant condition. Therefore, any convenient frame can be chosen for construction of the solution states $\left|\xi_\Psi^{(n)}\right\rangle$.

In order to construct these states, one must first make momentum expansions of the quark fields that are consistent with (2.5) and (2.8). Using the Dirac representation of the gamma matrices, the quark fields can be expanded as follows:

$$\Psi = \sum_{\alpha s} C_\alpha \int \frac{d^3p}{p_0 \sqrt{2(2\pi)^3}} \begin{pmatrix} p^0 S_s \\ \vec{p} \cdot \vec{\sigma} S_s \end{pmatrix} \left(b_{\alpha s p} e^{-ip \cdot x} + \overline{b}_{\alpha s \overline{p}} e^{ip \cdot x}\right)$$

$$\chi = \sum_{\alpha s} C_\alpha \int \frac{d^3p}{2p_0^2 \sqrt{2(2\pi)^3}} \begin{pmatrix} p^0 S_s \\ -\vec{p} \cdot \vec{\sigma} S_s \end{pmatrix} \left(b_{\alpha s p} e^{-ip \cdot x} - \overline{b}_{\alpha s \overline{p}} e^{ip \cdot x}\right), \tag{2.13}$$

where $p^0 = |\vec{p}|$, $\overline{p}^\mu = p_\mu = (p^0, -\vec{p})$, and $C_\alpha$ and $S_s$ are color and spin unit vectors, respectively. It is easily verified that these fields satisfy the "massless" field equation (2.5) as well as $\Psi = i\partial\chi$, which is the conjugate of the potential field equation (2.3). It should be noted that no transformation is made from negative-energy quark destruction operators $\overline{b}_{\alpha s \overline{p}}$ to positive-energy anti-quark creation operators. The choice to use negative-energy quark operators is made to simplify later analyses, although it is also possible to perform the entire following analyses using antiquark operators rather than negative-energy quark operators.

The quark operators are quantized by requiring that they destroy a "bare" vacuum state $\left|\xi_\Psi^{(0)}\right\rangle$

$$b_{\alpha s p}\left|\xi_\Psi^{(0)}\right\rangle = \overline{b}_{\alpha s p}\left|\xi_\Psi^{(0)}\right\rangle = 0, \tag{2.14}$$

and by imposing the following anti-commutators

$$\{b_{\alpha s p}, b_{\alpha' s' p'}^+\} = \delta_{\alpha\alpha'}\delta_{ss'}\delta^3(\vec{p} - \vec{p}') \quad \text{and}$$

$$\{\overline{b}_{\alpha s p}, \overline{b}_{\alpha' s' p'}^+\} = \delta_{\alpha\alpha'}\delta_{ss'}\delta^3(\vec{p} - \vec{p}'), \tag{2.15}$$

with all other anti-commutators vanishing. This quantization condition leads to the field quantization of equation (2.8), as can be verified by direct calculation.

Since any frame can be used for identification of solution states that satisfy the Lorentz-invariant condition (2.12), there is no loss of generality by choosing the Hamiltonian frame $u = (1,0,0,0)$. In the Hamiltonian frame, the background gluon field has the following momentum expansion:

$$g\overline{A}_\mu^a = m \sum_{\alpha s} \overline{T}_\alpha^a \int \frac{d^3p}{2p_0} \left(p_\mu b_{\alpha s p}^+ \overline{b}_{\alpha s p} e^{2ip^0 t} + \overline{p}_\mu \overline{b}_{\alpha s p}^+ b_{\alpha s p} e^{-2ip^0 t}\right), \tag{2.16}$$

where $\overline{T}_\alpha^a = C_\alpha^T \overline{T}^a C_\alpha$. In this frame, the background gluon field is independent of spatial coordinates and only dependent on time. This means that in the quantity $\int d^4x g \overline{A}_\mu^a J^{\mu a}$, the background field can be taken out of the spatial integral. The remaining "charge" vector takes the form:

$$\int d^3x J^{\mu a} = Q^{\mu a} + \delta^{\mu i}\tilde{Q}^{ia}$$

$$Q^{\mu a} = \sum_{\alpha s} T^a_\alpha \int \frac{d^3p}{p^0} p^\mu \left(b^+_{\alpha sp} b_{\alpha sp} + \bar{b}^+_{\alpha s\bar{p}} \bar{b}_{\alpha s\bar{p}}\right)$$

$$\tilde{Q}^{ia} = i\sum_{\alpha ss'} T^a_\alpha \int \frac{d^3p}{p^0} S^T_{s'} \varepsilon^{ijk} p^j \sigma^k S_s \left(\bar{b}^+_{\alpha s'p} b_{\alpha sp} e^{-2ip^0 t} - b^+_{\alpha s'p} \bar{b}_{\alpha sp} e^{2ip^0 t}\right), \quad (2.17)$$

where the above expressions are for the color-diagonal charges since those are the only ones that contribute to $\int d^4x g\bar{A}^a_\mu J^{\mu a}$.

Using (2.17), the background interaction term takes the form:

$$\int d^4x g\bar{A}^a_\mu J^{\mu a} = \left(\int dt g\bar{A}^a_\mu\right) Q^{\mu a} + \int dt \left(g\bar{A}^a_i \tilde{Q}^{ia}\right). \quad (2.18)$$

Because of the time dependence of the background field, the first term above vanishes for all terms within $g\bar{A}^a_\mu$ except those with $|\vec{p}|=0$. Thus for any states not involving $|\vec{p}|=0$ quarks in the Hamiltonian frame, the first term of (2.18) can be ignored. For these states, the following simpler condition is required to ensure that (2.12) is satisfied

$$\int dt \left(g\bar{A}^a_i \tilde{Q}^{ia}\right) \left|\xi^{(n)}_\Psi\right\rangle = \left\langle\xi^{(n)}_\Psi\right| \int dt \left(g\bar{A}^a_i \tilde{Q}^{ia}\right) = 0. \quad (2.19)$$

It can be directly verified that

$$\left[g\bar{A}^a_i, \tilde{Q}^{ia}\right] = 0 \quad (2.20)$$

since it is proportional to $\varepsilon^{ijk} p^i p^j$. As a result, the two relations in (2.19) are equivalent.

It should be noted that for states involving $|\vec{p}|=0$ quarks in the Hamiltonian frame, $\int dt g\bar{A}^a_i$ in the first term of (2.18) can still be taken to vanish by using a vanishing infrared mass in its calculation. Furthermore, since $\left[g\bar{A}^a_0, Q^{0a}\right] = 0$, the remaining part of that term will vanish for $|\vec{p}|=0$ states that satisfy $Q^{0a}\left|\xi^{(n)}_\Psi\right\rangle = 0$. This aside is important since the vacuum that will be proposed in (3.12) includes quark operators at every momentum value, including $|\vec{p}|=0$.

There are many classes of solutions to (2.19) and the other equation of motion (1.12). An explicit enumeration of all possible classes of solutions is outside the scope of this paper. Nonetheless, the notation $\left|\xi^{(n)}_\Psi\right\rangle$ will continue to refer to any possible solution of the equations of motion. In the present section, the focus will remain on identifying solution states whose symmetries, energy, and independence on the choice of $u$ make them good candidates for the vacuum. Due to (2.19) and (2.20), it follows that one class of solutions is defined by states satisfying either $g\bar{A}^a_i\left|\xi^{(n)}_\Psi\right\rangle = 0$ or $\tilde{Q}^{ia}\left|\xi^{(n)}_\Psi\right\rangle = 0$. Solutions in this class will now be identified.

Let us define the following composite operators:

$$P^+_{sp} = \prod_\alpha b^+_{\alpha sp} \bar{b}^+_{\alpha sp}$$

$$V_p^+ = \frac{1}{2\sqrt{2}} \prod_\alpha \left(b_{\alpha\uparrow p}^+ \bar{b}_{\alpha\downarrow p}^+ + \bar{b}_{\alpha\uparrow p}^+ b_{\alpha\downarrow p}^+\right)$$

$$F_p^+ = P_{\uparrow p}^+ P_{\downarrow p}^+ = V_p^+ V_p^+ = \prod_{\alpha s} b_{\alpha s p}^+ \bar{b}_{\alpha s p}^+ , \tag{2.21}$$

where the last operator is a "full" operator in the sense that in the present one-flavor approach, it contains every possible quark operator at a given momentum. As a result, it is not possible to have a state with a "full" operator and some other creation operator at the same momentum. It can be directly verified that the operators of (2.21) satisfy the following commutation relations:

$$\left[g\bar{A}_\mu^a, P_{sp}^+\right] = 0$$
$$\left[\widetilde{Q}^{i3}, V_p^+\right] = \left[\widetilde{Q}^{i8}, V_p^+\right] = 0$$
$$\left[g\bar{A}_\mu^a, F_p^+\right] = \left[\widetilde{Q}^{ia}, F_p^+\right] = 0. \tag{2.22}$$

Since both the background field and charge operator annihilate the bare vacuum, it follows that a class of solutions to (2.19) can be found by making different combinations of the operators of (2.21) acting on the bare vacuum. One restriction is that for any given value of $|\vec{p}|$, one cannot have both $P$ and $V$ operators, only $P$ or $V$ operators. This restriction will become clearer when discussing other solutions below.

Other states that solve the equations of motion are ones that feature only "positive-energy" quarks or only "negative-energy" quarks for a given value of $|\vec{p}|$. To see this, it is useful to write out the operator in (2.19) after performing the time integral. One has:

$$\int dt g\bar{A}_i^a \widetilde{Q}^{ia} = \pi m \sum_{a\alpha\beta s's''} \bar{T}_\alpha^a T_\beta^a \int \frac{d^3 p}{2p_0} \frac{d^3 q}{2q_0} p_i \varepsilon^{ijk} q^j S_{s''}^T \sigma^k S_{s'} \delta(p_0 - q_0) \times$$
$$\times \left(b_{\alpha s p}^+ \bar{b}_{\beta s'' q}^+ \bar{b}_{\alpha s p} b_{\beta s' q} + b_{\beta s'' q}^+ \bar{b}_{\alpha s p}^+ \bar{b}_{\beta s' q} b_{\alpha s p}\right) \tag{2.23}$$

Because of the $\delta(p_0 - q_0)$ factor, all of the quark operators in a given term must have the same value of $|\vec{p}|$. However, since each term involves both a $b$ and a $\bar{b}$ operator, states involving only $b^+$ operators or only $\bar{b}^+$ operators at a given value of $|\vec{p}|$ are annihilated by the operator $\int dt g\bar{A}_i^a \widetilde{Q}^{ia}$. Similar reasoning causes the restriction mentioned in the last paragraph on states made out of $P$ and $V$ operators. One notable solution state in the present class is the state $\prod_{\alpha s} \bar{b}_{\alpha s p}^+ |\xi_\Psi^{(0)}\rangle$ involving every possible "negative-energy" quark operator for a particular momentum.

One requirement for all solution states $|\xi_\Psi^{(n)}\rangle$, including ones not explicitly discussed above, is that they must involve symmetric sums or products over all fundamental color indices. As a result, the following relations hold:

$$\left\langle \xi_\Psi^{(n)} \left| \bar{D}_\mu^{ab} \bar{F}^{\mu\nu b} \right| \xi_\Psi^{(m)} \right\rangle = \left\langle \xi_\Psi^{(n)} \left| J^{\mu a} \right| \xi_\Psi^{(m)} \right\rangle = 0. \tag{2.24}$$

Note that each of the operators in the above matrix elements has an open adjoint color index. Since the only color matrices in the theory are the group generators, it follows that the matrix elements of (2.24) must be proportional to $\sum_\alpha T_\alpha^a$. Because the group generators of SU(3) are traceless, these matrix elements vanish. Not only does this mean that the final equation of motion (1.12) is satisfied for states with this kind of color symmetry, it also means that it is satisfied without any dependencies between quark and gluon-fluctuation sectors of the theory – every mixed quark-gluon-fluctuation term vanishes independently. This lack of interdependence in the equations of motion means that in the context of solution states, quark and gluon-fluctuation operators commute with each other. This justifies the earlier separation of quark and gluon-fluctuation operators into separate quantum states.

It should be noted that if the fermion mass matrix of (1.2) were not taken to be SU(3) symmetric (having the same mass value for each SU(3) index), then the solutions found here would not apply. Given an asymmetric mass matrix $M$, one could still define the background gluon field via (2.6), but the $M\overline{T}^a$ factor in that expression would no longer be traceless. This would mean that the background gluon field would no longer be traceless, so the trace of $\overline{D}_\mu^{ab}\overline{F}^{\mu\nu b}$ would not vanish, and equation of motion (1.12) would not be solved by states such as the ones presented here. For this reason, the diagonal background field formalism presented here cannot be used for the SU(2) symmetry in the electro-weak interaction, since the fermions in SU(2) doublets do not have the same mass. In addition, the present approach does not work for QED since the group generator is not traceless.

**The Hamiltonian, the vacuum, and stability**

Now that exact solutions states have been found, one would like to find the lowest-energy solution state and show that it is stable to all fluctuations. This is done in the following steps: First, the Hamiltonian in the presence of an operator background field is derived. Using that Hamiltonian, the lowest-energy solution state is found and identified as the vacuum $|0\rangle$. This vacuum is shown to have infinitely negative absolute energy and be stable to any fluctuations within the class of solution states $|\xi\rangle$. Finally, the vacuum is shown to be stable to fluctuations outside of the class $|\xi\rangle$, including all colored fluctuations.

Calculation of the Hamiltonian for a general value of the 4-velocity $u$ is complicated due to the fact that the quark fields within a background field factor may have different time coordinates from those not inside a background field factor. However, if one takes $u = (1,0,0,0)$, all fields in the Lagrangian have the same time coordinate, whether or not they are inside a background field. With this choice of $u$, quantum field theory can be reduced to quantum mechanics through the usual method of treating the time-dependent field at each point of space as a separate canonical variable. Since the equations of motion are independent of the choice made for $u$, one is free to choose $u = (1,0,0,0)$ in deriving the Hamiltonian. Setting $u$ to this value means that $u$ represents the same reference frame as the one chosen for calculation of the Hamiltonian. It is for this reason that $u = (1,0,0,0)$ has been referred to throughout as the "Hamiltonian frame".

Before calculating the Hamiltonian, one must decide which variables to use as the canonical variables. The usual choice for quarks is to take $\Psi$ and $\overline{\Psi}$ as the canonical variables. However, in this case since the background field in (2.7) depends explicitly on $\overline{\chi}$ rather than $\overline{\Psi}$, the expressions are greatly simplified if $\overline{\chi}$ is taken as a canonical variable instead of $\overline{\Psi}$. This choice does not present any new complications since the action is still only dependent on first time derivatives of $\overline{\chi}$ and no second time derivatives. As a result, the usual canonical formalism can be used throughout.

In order to consider the stability of solutions with respect to gluon fluctuations, a gluon fluctuation operator must be included in the complete gluon field used for calculation of the Hamiltonian. In other words, one can substitute

$$A_\mu^a = \overline{A}_\mu^a + \hat{A}_\mu^a \qquad (3.1)$$

into the action (1.1). The resulting Lagrangian in the Hamiltonian frame is given by:

$$L = L_\Psi + L_A + L_{\Psi A}$$

$$L_\Psi = \int d^3x \left[ \overline{\Psi} i \partial \Psi + g\overline{A}_\mu^a J^{\mu a} + \tfrac{1}{2} \partial_0 \overline{A}_i^a \partial_0 \overline{A}_i^a \right]$$

$$L_{\Psi A} = \int d^3x \left[ g\hat{A}_\mu^a J^{\mu a} + \hat{F}_{0i}^a \partial_0 \overline{A}_i^a - gf^{abc} \hat{F}^{\mu\nu a} \hat{A}_\mu^b \overline{A}_\nu^c - \tfrac{1}{2} g^2 f^{abc} f^{ade} \hat{A}^{\mu b} \overline{A}^{\nu c} \left( \hat{A}_\mu^d \overline{A}_\nu^e + \overline{A}_\mu^d \hat{A}_\nu^e \right) \right]$$

$$L_A = -\tfrac{1}{4} \int d^3x \hat{F}_{\mu\nu}^a \hat{F}^{\mu\nu a}, \qquad (3.2)$$

where $L_\Psi$ involves only quark operators, $L_A$ involves only gluon-fluctuation operators, and $L_{\Psi A}$ involves both. It should be noted that the commutator of (2.9) has been used to cancel the quark mass term in $L_\Psi$. In addition, the fact that the background field is diagonal and independent of spatial coordinates has been used to set $\overline{F}_{ij}^a = 0$ and $\overline{F}_{0i}^a = \partial_0 \overline{A}_i^a$.

In calculating the Hamiltonian from the Lagrangian of (3.2), a subtle point is the calculation of contributions to quark canonical momentum coming from the background gluon field. For example, using (2.7) one has:

$$\frac{\delta(g\overline{A}_\mu^a(t))}{\delta(\partial_0 \overline{\chi}(t,\vec{y}))} = i\delta_{\mu 0} \gamma^0 M \overline{T}^a \Psi(t,\vec{y}). \qquad (3.3)$$

Adding up canonical momentum contributions analogous to $\dot{q}p$ from every point of space $\vec{y}$ within the field $gA_0^a(t)$, one has:

$$\int d^3y \partial_0 \overline{\chi}(t,\vec{y}) \frac{\delta(g\overline{A}_\mu^a(t))}{\delta(\partial_0 \overline{\chi}(t,\vec{y}))} = \delta_{\mu 0} g\overline{A}_0^a(t). \qquad (3.4)$$

Similarly, the $\dot{q}p$ canonical momentum contributions coming from the background field strength term become

$$\int d^3y \left[ \partial_0 \overline{\chi}(t,\vec{y}) \frac{\delta(\partial_0 \overline{A}_i^a(t))}{\delta(\partial_0 \overline{\chi}(t,\vec{y}))} + \frac{\delta(\partial_0 \overline{A}_i^a(t))}{\delta(\partial_0 \Psi(t,\vec{y}))} \partial_0 \Psi(t,\vec{y}) \right] = \partial_0 \overline{A}_i^a(t). \qquad (3.5)$$

Using expressions like those above along with standard canonical methods, it is straightforward to see that matrix elements of the Hamiltonian take the form:

$$\langle \xi' | H | \xi \rangle = \langle \xi' | (H_\Psi + H_A + H_{\Psi A}) | \xi \rangle$$

$$H_\Psi = -\int d^3x \overline{\Psi} i \gamma^i \partial_i \Psi + \tfrac{1}{2} \left( \int d^3x \right) (\partial_0 \overline{A}_i^a(t))^2$$

$$H_{\Psi A} = \tfrac{1}{2} \sum_{abc} f^{abc} f^{abc} \int d^3x \left[ ((g\overline{A}_0^b)^2 + (g\overline{A}_j^b)^2)(\hat{A}_i^c)^2 - (g\overline{A}_j^b)^2 (\hat{A}_0^c)^2 - g\overline{A}_i^b g\overline{A}_j^b \hat{A}_i^c \hat{A}_j^c \right]$$

$$H_A = \tfrac{1}{2} \int d^3x \left( \hat{F}_{0i}^a \hat{F}_{0i}^a + \tfrac{1}{2} \hat{F}_{ij}^a \hat{F}_{ij}^a \right), \qquad (3.6)$$

where (2.12) and (2.24) have been used. It should be noted that since the background gluon field is independent of spatial coordinates, the spatial integral in the background "electric" field energy term of the quark Hamiltonian becomes an infinite volume factor $\left(\int d^3x\right)$. It will be seen shortly that this factor causes the energy of the proposed vacuum to be infinitely more negative than that of vacuum candidates built on the Dirac Sea.

Now that the Hamiltonian has been constructed, it is possible to determine the energy associated with any of the quark solution states $\left|\xi_\Psi^{(n)}\right\rangle$ that were explicitly defined in the last section. Moreover, the state with the lowest possible energy can also be identified. To do that, it is easiest to focus first on $H_\Psi$ and then to consider the rest of the Hamiltonian afterward. Starting with the background electric field energy term of $H_\Psi$, it is useful to use (2.16) to write down the momentum expansion of the background electric field strength in the Hamiltonian frame:

$$\partial_0 \overline{A}_i^a = i\frac{m}{g} \sum_{\alpha s} \overline{T}_\alpha^a \int d^3 p\, p_i \left(b_{\alpha sp}^+ \overline{b}_{\alpha sp} e^{2ip^0 t} + \overline{b}_{\alpha sp}^+ b_{\alpha sp} e^{-2ip^0 t}\right). \tag{3.8}$$

This field strength is manifestly anti-Hermitian. This means that the square of the background field strength has either negative or zero expectation values for any quark solution state. The vacuum state should be one that maximizes this negative energy.

It is useful first to eliminate some vacuum candidates. One finds:
$$\left[\partial_0 \overline{A}_i^a, P_{sp}^+\right] = 0$$
$$\left[\partial_0 \overline{A}_i^a, F_p^+\right] = 0 \tag{3.9}$$

As a result, adding a $P$ or $F$ operator to a state does not add any negative electric energy. Moreover, the fact that those operators fill up possible slots for other composite quark operators that do lead to negative energy means that they are not good candidates for the vacuum.

Let us now calculate the quark energy of two other candidate states. It can be verified that:

$$\left\langle\xi_\Psi^{(n)}\left|H_\Psi \prod_{\alpha s}\overline{b}_{\alpha sp}^+\right|\xi_\Psi^{(0)}\right\rangle = \left[-6p_0 - \frac{18m^2 p_0^2}{g^2}\left(\int d^3x\right)\right]\left\langle\xi_\Psi^{(n)}\left|\prod_{\alpha s}\overline{b}_{\alpha sp}^+\right|\xi_\Psi^{(0)}\right\rangle \tag{3.10}$$

$$\left\langle\xi_\Psi^{(n)}\left|H_\Psi V_p^+\right|\xi_\Psi^{(0)}\right\rangle = -\frac{36m^2 p_0^2}{g^2}\left(\int d^3x\right)\left\langle\xi_\Psi^{(n)}\left|V_p^+\right|\xi_\Psi^{(0)}\right\rangle \tag{3.11}$$

In each of the above equations, when $H_\Psi$ acts on the solution state to its right, it returns the same state as well as some states that are not solution states. When the non-solution states created by $H_\Psi$ are acted on the left by any solution state $\left\langle\xi_\Psi^{(n)}\right|$, they vanish. It should be emphasized that when $H_\Psi$ acts on a solution state, it does not return any new solution states. As a result, equations (3.10) and (3.11) are like eigen-equations in the restricted space of solution states.

In equation (3.10) the first term in square brackets is the kinetic energy and the second is the background electric field energy. The factor of 18 is comprised of a factor of 9 from $\text{Tr}\left(\overline{T}_\alpha^a \overline{T}_\alpha^a\right)$ and a factor of 2 from spins. The state in (3.11) gets an additional factor of 2 due to an additional way that $H_\Psi$ can return the same state. For Lorentz-invariant plane wave quantization, the spatial volume factor $\left(\int d^3x\right)$ can be taken to be infinite. As a result, the background electric energy at a given momentum completely overwhelms the kinetic energy. This means that the state in (3.11) has infinitely more negative energy than the state in (3.10).

Consequently, the following quark vacuum is proposed:
$$|0_\Psi\rangle \equiv \prod_p V_p^+ |\xi_\Psi^{(0)}\rangle, \qquad (3.12)$$
since this state maximizes the negative quark energy at every value of the momentum. Although the vacuum was defined in the Hamiltonian frame, the vacuum is frame-independent due to the product over all momenta in (3.12). In particular, the vacuum state is independent of the 4-velocity parameter $u$ in the background field.

Using the same reasoning as that used for equation (2.24), it is apparent that the vacuum satisfies the condition
$$\langle 0_\Psi | \overline{A}_\mu^a | 0_\Psi \rangle = 0, \qquad (3.13)$$
no matter what form is taken by the gluon fluctuation sector of the vacuum. Since the vacuum state is a composite $|0\rangle = |0_\Psi\rangle |0_A\rangle$ of the quark vacuum and some gluon-fluctuation vacuum state, the Lorentz condition of (1.13) is satisfied.

Turning to the interaction Hamiltonian $H_{\Psi A}$, one can now consider the stability of the proposed quark vacuum with respect to gluon fluctuations. One finds that the quark vacuum expectation value of the interaction Hamiltonian takes the form:
$$\langle 0_\Psi | H_{\Psi A} | 0_\Psi \rangle = \tfrac{1}{2} \left( \sum_p m^2 \right) \int d^3x \sum_{a \neq 3,8} \left[ 9(\hat{A}_0^a)^2 + 3(\hat{A}_i^a)^2 \right]. \qquad (3.14)$$
In the context of the quark vacuum, the interaction Hamiltonian gives infinite mass proportional to $m\sqrt{\sum_p 1}$ to all of the non-diagonal gluon fluctuations. Because of this infinite mass, all of the gluon-fluctuation non-Abelian interactions effectively vanish, and the effective gluon-fluctuation Hamiltonian becomes a free Hamiltonian in which the non-diagonal gluons are massive. In the $\hat{A}_0^a = \partial_i \hat{A}_i^a = 0$ gauge one finds:
$$\langle 0_\Psi | (H_g + H_{qg}) | 0_\Psi \rangle \cong \int d^3x \left[ \sum_a ((\hat{E}_i^a)^2 + (\hat{B}_i^a)^2) + \tfrac{3}{2} \left( \sum_p m^2 \right) \sum_{a \neq 3,8} (\hat{A}_i^a)^2 \right], \qquad (3.15)$$
where $\hat{E}_i^a = \partial_i \hat{A}_0^a$ and $\hat{B}_i^a = \varepsilon_{ijk} \partial_j \hat{A}_k^a$. It is clear that the above quantity is positive-definite. It follows that gluon fluctuations can only add to vacuum energy and that therefore the vacuum of (3.12) is stable to all gluon fluctuations.

In addition to being stable to gluon fluctuations, the vacuum is also stable to quark fluctuations. The argument is as follows: First, the quark vacuum was constructed to have lower energy than any other plane-wave quark solution state $|\xi_\Psi^{(n)}\rangle$ and is therefore stable to fluctuations to other solution states as defined in this paper. Next, it will be shown that the vacuum is stable to fluctuations to any quark plane-wave states that are not solution states.

To consider plane-wave states $|\phi\rangle$ outside of the class $|\xi\rangle$ of solution states that satisfy (1.10)-(1.12), one must use standard techniques that do not involve an operator background field. Using these standard techniques, the largest contributions to vacuum energy come from the large momentum $p \to \infty$ contributions. In this sector, quark-gluon couplings vanish and it is valid to treat quark and gluon sectors of the theory independently. In this limit, the lowest-energy quark state for a given momentum would be the Dirac-sea state, which has energy $-6\sqrt{|\vec{p}|^2 + m^2}$, due to the kinetic energy of 3 colors and 2 spins of negative-energy quarks at a given momentum. The finite negative energy of this state should be compared to the negative energy of (3.11) that is infinite as a result of the infinite spatial volume factor $\left(\int d^3x\right)$. Because of this difference in energies, any fluctuation to a plane-wave state $|\phi\rangle$ outside of the class $|\xi\rangle$ of plane-wave solution states is infinitely suppressed. In other words, the proposed vacuum is stable to fluctuations to any other Lorentz-invariant state.

**4. Confinement and Hadrons**

Until now the focus has been on finding vacuum solutions. As a result, only Lorentz-invariant quantization and configurations have been considered. Since Lorentz-invariant quantization takes place inside an infinite spatial volume, the spatial volume factor $\left(\int d^3x\right)$ on the background electric field energy has been taken to be infinite. For an excitation above the vacuum to have finite energy, the spatial volume factor $\left(\int d^3x\right)$ relevant for that excitation must be finite. It follows that finite-energy solutions must involve fields that are contained within finite spatial volumes. However, even inside finite volumes, it will be shown that colored quark solutions have infinite energy relative to the vacuum. Thus, the proposed vacuum exhibits color confinement. Later in the section, it is shown that colorless cubical finite-energy solutions exist that have the quantum numbers of hadrons. The masses and radii of these cubical "hadrons" can be self-consistently determined through energy minimization.

To consider field configurations contained within a finite region of space, one must separately quantize the fields inside the region and those outside the region. One can assume that the fields outside the region are equivalent to the vacuum proposed in the last section. Assuming all fields vanish or are periodic on the surface of the region, the creation and destruction operators associated with the interior fields can be analyzed independently of and considered to commute with the operators outside the region.

Now let us consider a colored state $|\phi\rangle$ inside some finite region of space. Such a state would feature

$$\langle \phi | J^{\mu a} | \phi \rangle \neq 0 \ , \tag{4.1}$$

so the background field method presented in this paper would not be valid. To determine the energy of such a state, one would have to revert to more standard perturbative techniques. Just as in the discussion at the end of the last section, the lower limit of the energy would be that of the Dirac Sea within the region, since colored configurations "on top of" the Dirac Sea would just increase the energy. Therefore, the quark energy of a colored quark state would be

$$E_D > -6 \sum_p \sqrt{|\vec{p}|^2 + m^2} \ , \tag{4.2}$$

where $\vec{p}$ would run over all values consistent with the fields vanishing or being periodic on the surface of the region.

On the other hand, the energy of that same region produced by the vacuum proposed in this paper is (see (3.11)):

$$E_0 = -\frac{36m^2}{g^2}\left(\int d^3x\right)\sum_p |\vec{p}|^2 . \qquad (4.3)$$

Even for very small regions $\left(\int d^3x\right)$, $E_0$ is infinitely more negative than $E_D$ due to the fact that it has a second-degree momentum divergence rather than a first-degree divergence. This argument holds for regions all the way down to those that have inverse dimension equal to Planck-scale momentum cutoffs. It follows that colored quark states, even in finite-sized regions, have infinite energy relative to the vacuum. In other words, the proposed vacuum exhibits colored quark confinement.

Next, let us discuss finite-energy colorless "hadron" solutions confined inside a finite region. A correct discussion of hadrons that arise from the background field method of this paper would require canonical quantization of quark fields inside a sphere. This topic is somewhat complicated and will be addressed in a later paper. However, one can understand many of the important features of hadrons, including how their masses and radii are self-consistently determined, by considering fields quanitzed inside a cube with side $2L$. In this "cubical quantization", the quark momentum must take only the values $\vec{p} = \frac{\pi}{L}(n,m,l)$ where $n$, $m$, and $l$ are integers. This is to ensure that the fields are periodic on the surface of the cube. In addition to these general allowed values of momentum, it is useful to define the following specific momenta

$$\vec{q} = \frac{\pi}{L}(1,0,0)$$
$$\vec{q}' = \frac{\pi}{L}(0,1,0)$$
$$\vec{q}'' = \frac{\pi}{L}(0,0,1) . \qquad (4.4)$$

All of the above momenta (and their opposites, $-\vec{q}$, etc.) have the same magnitude, but different directions, so they represent distinct quantum states.

In the context of cubical quantization, it is useful to introduce flavor and define the following composite operators:

$$V^+_{fp} = \frac{1}{2\sqrt{2}}\prod_\alpha \left(b^+_{\alpha\uparrow fp}\bar{b}^+_{\alpha\downarrow fp} + \bar{b}^+_{\alpha\uparrow fp}b^+_{\alpha\downarrow fp}\right)$$

$$F^+_{fp} = \prod_\alpha b^+_{\alpha\uparrow fp}\bar{b}^+_{\alpha\downarrow fp}b^+_{\alpha\uparrow \bar{f}\bar{p}}\bar{b}^+_{\alpha\downarrow \bar{f}\bar{p}} = V^+_{fp}V^+_{\bar{f}\bar{p}}$$

$$B^+_{ss's''ff'f''p} = \frac{1}{\sqrt{6}}\sum_{\alpha\beta\gamma}\varepsilon_{\alpha\beta\gamma}b^+_{\alpha s fp}b^+_{\beta s'f'p}b^+_{\gamma s''f''p}$$

$$\bar{B}^+_{ss's''ff'f''\bar{p}} = \frac{1}{\sqrt{6}}\sum_{\alpha\beta\gamma}\varepsilon_{\alpha\beta\gamma}\bar{b}^+_{\alpha s f\bar{p}}\bar{b}^+_{\beta s'f'\bar{p}}\bar{b}^+_{\gamma s''f''\bar{p}}$$

$$M^+_{ss'ff'p} = \frac{1}{\sqrt{3}}\sum_\alpha b^+_{\alpha s fp}\bar{b}_{\beta s'f'\bar{p}} , \qquad (4.5)$$

where *f* indices denote flavor, and the index $\bar{s}$ denotes spin in the opposite direction from *s*. For reference, the above operators will be called "vacuum", "full", "baryon", "anti-baryon", and "meson" operators, respectively. Given this naming, one can see the correspondence between the negative-energy quark representation used so far and the usual anti-quark representation. For example, an anti-baryon creation operator is comprised of three negative-energy quark destruction operators. As usual, these are equivalent to three positive-energy anti-quark creation operators.

It is also convenient to define the flavor number operators as:

$$N_f \equiv \int d^3 x \Psi_f^+ \Psi_f = \sum_{\alpha s} \int d^3 p \left( b_{\alpha s f p}^+ b_{\alpha s f p}^+ + \bar{b}_{\alpha s f p}^+ \bar{b}_{\alpha s f p}^+ \right). \quad (4.6)$$

With these definitions, it can be seen that $F_{fp}^+$ has twice the flavor number of $V_{fp}^+$, or the same flavor number as $V_{fp}^+ V_{fp}^+$. Since flavor numbers measure both baryon number and electric charge, keeping track of flavor numbers is important in identifying hadron solutions.

The quark vacuum of the theory is given by the flavor generalization of (3.12):

$$\left| 0_\Psi \right\rangle \equiv \prod_{fp} V_{fp}^+ \left| \xi_\Psi^{(0)} \right\rangle, \quad (4.7)$$

with the definition of allowable momenta *p* restricted by the cubical quantization. In addition to the vacuum, let us also define the following state relative to the specific momenta *q* defined in (4.4):

$$\left| h_q \right\rangle \equiv V_{fq}^+ V_{f\bar{q}}^+ F_{fq'}^+ F_{fq''}^+ \prod_{p^0 \neq q^0} V_{fp}^+ \left| \xi_\Psi^{(0)} \right\rangle. \quad (4.8)$$

Since $\tilde{Q}^{ia} \left| h_q \right\rangle = 0$, this state is a solution to the equations of motion. It can also be seen that $\left| h_q \right\rangle$ has the same flavor number as the quark vacuum of (4.7). It will now be shown that this state is needed as a basis for the construction of hadron states.

Cubical "hadron" states are created by acting on $\left| h_q \right\rangle$ with baryon, anti-baryon, or meson operators. In particular, one has

$$\left| B_{ss's''ff'f''q} \right\rangle \equiv B_{ss's''ff'f''q}^+ \left| h_q \right\rangle$$
$$\left| \bar{B}_{ss's''ff'f''q} \right\rangle \equiv \bar{B}_{ss's''ff'f''q}^+ \left| h_q \right\rangle$$
$$\left| M_{ss'ff'q} \right\rangle \equiv M_{ss'ff'q}^+ \left| h_q \right\rangle \quad (4.9)$$

Even though $\left[ \tilde{Q}^{ia}, B_{ss's''ff'f''q}^+ \right] \neq 0$, with similar relations for the other hadron operators, the above states all satisfy (2.19) and hence solve the equations of motion. This can be directly verified using the flavor generalization of equation (2.23). Essentially, that operator causes hadron states to vanish in one of four ways: making $\delta(p_0 - q_0)$ vanish, creating a term with dependence $\varepsilon^{ijk} q^i q^j$, creating a term with two identical $b^+$ operators, or annihilating the bare vacuum.

It should also be noted that $B_{ss's''ff'f''q}^+ \left| 0_\Psi \right\rangle$ is not a solution state since it would not vanish when the $\tilde{Q}^{ia}$ in (2.23) acted on the baryon operator and the $g\bar{A}_i^a$ in the same term acted on one of the vacuum operators such as $V_{q'}^+$ that have quark operators with the same momentum magnitude, but different direction from those in the baryon operator. It is to avoid this problem that $|h_q\rangle$ features the replacement of four "vacuum" operators with two "full" operators. Similar arguments hold for the other hadrons.

Now that cubical hadron solutions have been found, it is interesting to calculate their energy and see that the minimization of that energy leads to self-consistently determined hadron masses and radii. Looking at the baryon solution, one finds that its energy relative to the vacuum of (4.7) is given by:

$$E_B(L) = \langle B_{f\!f'\!f''q} |H| B_{f\!f'\!f''q}\rangle - \langle 0_\Psi |H| 0_\Psi\rangle = 3\left(\frac{\pi}{L}\right) + \frac{(2L)^3}{g^2}\left(\frac{\pi}{L}\right)^2\left[9\left(m_f^2 + m_{f'}^2 + m_{f''}^2\right) + 144\sum_g m_g^2\right],$$
(4.10)

where spin indices on the baryon state have been suppressed, since in this cubical quantization, the spin of the baryon does not affect its mass. The first term on the right is just the kinetic energy $3q_0$ of the quark operators in the baryon operator (neither the quark vacuum nor the state $|h_q\rangle$ have any kinetic energy). The second term on the right involves the difference in background electric field energy between the baryon and the vacuum state and is proportional both to $q_0^2$ and to the hadron volume $(2L)^3$. The first term in the square brackets is the difference in energy between $B^+_{f\!f'\!f''q}\prod_f V^+_{fq}$ and $\prod_f V^+_{fq}$, while the second term in square brackets is the difference between $\prod_f F^+_{fq'}F^+_{fq''}$ and $\prod_f V^+_{fq'}V^+_{fq''}V^+_{fq'}V^+_{fq''}$. As stated in the last paragraph, the "full" operators are needed in these hadron states due to the fact that $q_0 = q'_0 = q''_0$, but their respective momenta are in different directions.

Given the baryon energy of (4.10), one may wonder what is the appropriate value to take for $L$, the size of the baryon. Since the two terms in (4.10) have different dependence on $L$, there is a unique $L$ that can be found by minimizing the energy. This baryon size and the resulting baryon energy/mass are given by:

$$L_B = \frac{g}{2\sqrt{6\pi\left(m_0^2 + m_f^2 + m_{f'}^2 + m_{f''}^2\right)}}$$

$$m_B \equiv E_B(L_B) = \frac{12}{g}\sqrt{6\pi^3\left(m_0^2 + m_f^2 + m_{f'}^2 + m_{f''}^2\right)},$$
(4.11)

where

$$m_0^2 \equiv 16\sum_g m_g^2$$
(4.12)

is a scale set by the sum over all possible flavors of quarks. Equation (4.11) says that in this model, stable baryon masses are completely determined by current quark masses in the Lagrangian and the strong coupling constant. The same statement can be made for the other "hadron" solutions of (4.9).

It is interesting to look at the predictions of this model for mass separations of baryons with increasing numbers of strange quarks. Let $m_1$, $m_2$, and $m_3$ represent down, up, and strange quarks, respectively, and consider the case where $m_0^2 \gg m_3^2 \gg m_2^2 \cong m_1^2$. In this case, a Taylor expansion of (4.11) would lead to the conclusion that:

$$m_\Omega - m_\Xi \cong m_\Xi - m_\Sigma \cong m_\Sigma - m_\Delta \cong \frac{6\sqrt{6\pi^3} m_3^2}{g m_0}, \qquad (4.13)$$

where $m_\Omega$, $m_\Xi$, $m_\Sigma$, and $m_\Delta$ denote baryons with 3, 2, 1, and 0 strange quarks, respectively, with all other quarks either up or down. Equation (4.13) states that this model features equal spacing of cubical baryon masses of increasing strangeness. This kind of equal spacing is observed in real baryons and originally led to the correct prediction for the Ω mass. However, there are two major problems with using the cubical baryon expression of (4.11) to attempt to reproduce actual baryon masses. Namely, it results in the same mass for both spin ½ and spin 3/2 baryons, and the value of $m_0$ in (4.12) is much too large given any reasonable estimate for top quark mass.

    Both of these problems are at least partially resolved when one moves from quantization inside a finite cube to quantization inside a finite sphere. The origin of spin-dependent masses in spherical quantization will not be discussed here, but a brief discussion of the $m_0$ problem will be made. Recall that the $m_0$ contribution in (4.11) came about due to allowed momenta $\vec{q}'$ and $\vec{q}''$ that had the same magnitude as the quark momenta in the hadron operators, but had different directions. In spherical quantization, in the $l = 0$ (no orbital angular momentum) spherical ground state, there is only one "momentum" state, so if a baryon solution could be made completely out of $l = 0$ quarks, the equivalent expression to (4.12) would feature $m_0 = 0$. However, in spherical quantization, the background field and charge vector are not diagonal in $l$, so the lowest-energy solution state involves a large contribution from $l = 0$ quarks but also small contributions from higher angular momentum quarks. These higher angular momentum contributions feature different angular momentum states with the same magnitude of "momentum", so they do introduce $m_0$ contributions. However, since the mix of higher angular momentum in the ground state is small, the effective $m_0$ from an expression like that of (4.12) is much smaller than in cubical quantization. These issues will be addressed much more fully in a coming paper involving spherical quantization of the present model.

**Summary and Future Work**

    New exact quantum solutions to the QCD equations of motion have been found. These solutions have no classical analog since they rely upon a background gluon field that does not commute with quark fields. This background gluon field is a Lorentz-covariant integral of quark fields whose functional form is restricted by the fact that it leads to exact extrema of the quantized QCD action. There are several classes of solution states associated with this background field. The lowest-energy state has been found and shown to be stable to all fluctuations. This "vacuum" state exhibits colored quark confinement and permits finite-energy hadron solutions as local minima of the energy.

    Exact solutions to quantum field theories are interesting in their own right. However, for the model presented here to be a true description of the strong interaction, a number of issues must be addressed. Given the phenomenological success of perturbative QCD, it would be useful to show that a pertubative QCD description is valid within the context of the present background field approach for short enough distances, deep inside some confined region of non-vacuum. There is no a priori reason perturbative QCD should not apply in this regime, but a more thorough examination is warranted.

One direct test of the approach will be whether it can reproduce the spectrum of hadron masses, given just current quark masses and the strong coupling constant as parameters. There was some discussion in the paper about how quantizing the theory inside a sphere will directionally improve agreement with key phenomenological observations, such as the Gell-Mann-Okubo mass formula. In addition, chiral symmetry breaking must be explicitly addressed and an explanation given for difference in masses of hadrons that have the same quark content but different angular momentum. These issues as well as a myriad of other connections to experimental data would all be resolved if a connection was made between the exact solutions presented here and the highly successful constituent quark models that have been developed [8-12]. These topics will be addressed in a coming paper.

**Acknowledgements**
I would like to thank Professor M. Suzuki, Professor U. Heinz, Dr. P. Lammert, and Dr. M. Thoma for invaluable feedback and advice in constructing this paper.